\documentstyle[12pt]{article}
\newcommand{\ox}{\otimes}
\newcommand{\w}{{\bf W}}
\newcommand{\wi}{\w_i}
\newcommand{\wj}{\w_j}
\newcommand{\tw}{\tilde{\bf W}}
\newcommand{\twi}{\tw_i}
\newcommand{\twj}{\tw_j}
\newcommand{\wo}{\w_0}
\newcommand{\z}{\w^{\prime}}
\newcommand{\zi}{\z_i}
\newcommand{\zj}{\z_j}
\newcommand{\kij}{k_{ij}}
\newcommand{\kji}{k_{ji}}
\newcommand{\koi}{k_{0i}}
\newcommand{\koj}{k_{0j}}
\newcommand{\kio}{k_{i0}}
\newcommand{\kii}{k_{ii}}
\newcommand{\tkij}{\tilde{k}_{ij}}
\newcommand{\tkji}{\tilde{k}_{ji}}
\newcommand{\tkoi}{\tilde{k}_{0i}}

\newcommand{\na}{N_{\alpha \w}}
\newcommand{\tna}{N_{{\tilde{\alpha}} \w}}
\newcommand{\nta}{N_{\alpha\tilde{\w}}}
\newcommand{\qa}{Q_{\alpha \w}}
\newcommand{\ai}{\alpha_i}
\newcommand{\aj}{\alpha_j}

\newcommand{\taj}{\tilde{\alpha}_j}
\newcommand{\h}{\frac{1}{2}}
\newcommand{\g}{\hspace{0.2 cm}}
\newcommand{\aw}{{\overline {\alpha \w}}}

\newcommand{\atw}{{\overline{\alpha\tilde{\w}}}}
\newcommand{\mod}{\mbox{\hspace{1 cm} mod(1)}}
\newcommand{\be}{\begin{equation}}
\newcommand{\ee}{\end{equation}}
\newcommand{\ba}{\begin{eqnarray}}
\newcommand{\ea}{\end{eqnarray}}
\newcommand{\ccc}{ {\mbox{\hspace{0.2 cm}}}\left| {\mbox{\hspace{0.2 cm}}} }
\newcommand{\lll}{\left[ {\mbox{\hspace{0.2 cm}}} }
\newcommand{\rrr}{{\mbox{\hspace{0.2 cm}}}\right.\right] }
\newcommand{\com}{ \mbox{,\hspace{0.1 cm}}}

\newcommand{\gen}{\overline{\mbox{gen}}}
\title{Classification of (2,2) Compactifications
       in Fermionic Strings}
\author{S.~A.~Abel$^{1}$ and C.~M.~A~Scheich$^{2}$
\vspace{0.3cm}
\\$^{1}$Rutherford Appleton Laboratory
\\Chilton, Didcot
\\Oxon OX11 0QX
\\title{Classification of (2,2) Compactifications
       in Fermionic Strings}
\author{S.~A.~Abel$^{1}$ and C.~M.~A~Scheich$^{2}$
\vspace{0.3cm}
\\$^{1}$Rutherford Appleton Laboratory
\\Chilton, Didcot
\\Oxon OX11 0QX
\\England
\\ \vspace{0.3cm}
\\$^{2}$Theoretical Physics Department
\\Keble Road
\\Oxford OX1 3NP
\\England}

\begin{document}

\small

\maketitle

\title{Abstract}

\begin{abstract}

We present a general scheme for generating $(2,2)$ symmetric
fermionic string models and classify the models
in $D=8$ and $D=6$ space-time dimensions with one twist.
It is pointed out that they allow the geometrical interpretation
as generalised torus compactifications. Their relation
to other compactifications is discussed and the
overlap with orbifolds is determined.

\end{abstract}

\pagebreak

Since string theory is an appealing candidate for the unification of
gravity with gauge interactions, an extensive knowledge of the
possible string theories in four dimensions is of wide interest.
The most promising candidate so far is the heterotic string, but
its construction can only be considered to be unique in $D=10$
space-time dimensions (if we choose the more promising version with
gauge group $E_8\otimes E'_8$). Any further reduction of space-time
dimensions leads at first glance to an embarrassingly large number of
models. This is especially true in the case of $D=4$.
Further studies however, have revealed that there are a suprisingly
small number of (potentially phenomenological viable) models.
Let us review the present situation for three large classes of models:

Compactifications on Calabi-Yau manifolds have been partially
classified in Ref.\cite{candelas}. Many of the models are related and
there are only a few three generation models \cite{tian,schimmi}.
Of special interest here is the subclass of models
that may also be realised by means of minimal $N=2$ superconformal
models \cite{G88}. They belong to special Calabi-Yau manifolds
with fixed moduli and therefore allow the extraction all
information relevant to their phenomenology \cite{wir}.
In addition, four
dimensional string theories may be constructed from superconformal
theories directly \cite{kazama-suzuki}. However only a little is known
about them.

Orbifolds are an especially interesting class of compactified string models.
Their phenomenological implications have been studied in great detail
\cite{ibanez}. $Z_N\otimes Z_M$ orbifolds seem to correspond to some
of the $(2,2)$ compactifications by $N=2$ minimal superconformal models
\cite{FIQ89}, whilst $Z_N$ orbifolds do not.

The last large class of models uses free fermions to construct models
directly in four dimensions, and the task of finding a
phenomenologically viable model has received considerable attention
\cite{nano}. They again belong to compactifications with fixed moduli.
This class has been compared to orbifolds in Ref.\cite{PS89}
by using theta-function identities. We will confirm their result
from another point of view.

Since the fermionic construction is very well suited for model building,
one should try to clarify its relation to the other
classes of compactification which to date remains rather obscure.
In order to do this we shall attempt to classify fermionic string models,
and to compare them to other models systematically.
We begin in this letter with the case of $D=8$ and $D=6$ and
postpone the study of $D=4$ to Ref.\cite{inprogress}.
This is because in these higher dimensions the models
are reduced drastically in number, but still show
the main features of their respective compactifications.

We furthermore restrict ourselves to symmetric $(2,2)$ models with
the maximal gauge group $g\otimes E_{D/2+4}\otimes E'_8$ where $g$ is a
model dependent gauge group of rank 2,5,8 for $D$ = 8,6,4 respectively
(for details see \cite{LT89}).
Further breaking of the gauge group by embeddings of twists should then
work in the usual way, and will not spoil the relevance of the classification.
Nonabelian embeddings that lower the rank are also not
considered. They correspond to using real fermions in the fermionic
string formulation. In addition, restricting the discussion to
left-right symmetric models also allows a direct interpretation
as a compactification.

We begin with a general discussion of the construction of $(2,2)$
fermionic strings that possess left-right symmetry.
We shall construct the models using the formulation of Ref.\cite{KLT87}
with only complex fermions. The internal fermions then have phases
associated with them $a_r,b_r,c_r$; $r=1,\cdots 5-D/2$ which come
in triplets for left and right movers and fulfill the constraint
\begin{equation}
a_r+b_r+c_r \g\in\g
\left\{ 0,\frac{1}{2}\right\} \mod .
\end{equation}
This constraint follows from the periodicity or antiperiodicity
of the superpartner of the stress-energy tensor. As shown in
Ref.\cite{KLT87} one derives the $N=1$ superconformal algebra
on the world-sheet.
The extension to $N=2$ \cite{banks} is rather cumbersome, and it is not
possible to realise the $N=2$ algebra for complex fermions in a closed form.
In general only matrix elements of the conformal fields are known.
The exception to this statement is the case where one complex fermion
is only ever periodic or antiperiodic in its boundary condition. It is
then possible to immediately form an $N=2$ superconformal algebra
\ba
T(z)   & = & \frac{1}{2}\sum_{i=1}^3:(:{\overline\psi}_i\partial\psi_i:
                                     +:\psi_i\partial{\overline\psi}_i:)
                                                 \nonumber\\
G^+(z) & = & -i \psi_1
           (:{\overline\psi}_3 \psi_3:
           -i:{\overline\psi}_2 \psi_2:)
                                                 \nonumber\\
G^-(z) & = & -i {\overline\psi}_1
           (:{\overline\psi}_3 \psi_3:
           +i:{\overline\psi}_2 \psi_2:)
                                                 \nonumber\\
J(z)   & = & :{\overline\psi}_1\psi_1:
\label{eq:fermions}
\ea
in a complex notation related to the real one by
$\psi=\frac{1}{\sqrt{2}}(\psi^1+i\psi^2)$,
$X=\frac{1}{\sqrt{2}}(X^1+iX^2)$.
Furthermore the space-time supersymmetry generator may only be realised
in a closed in the simplest of cases.
In order to compare models with only free fermions to
any of the models with
internal manifold coordinates we can simply bosonise two of the complex
fermions into one complex boson for each triplet (here we also require
left-right symmetry);
\be
\label{eq:bosonise}
\psi_i=:e^{iH}:
\ee
With this bosonisation we get immediately a torus compactification.
But this torus with its coordinates $H_i$ does not always
have the usual compactified former ten dimensional
coordinates as such. This is already obvious from the fact that
supersymmetry requires that a compactified coordinate
$X_i$ and its supersymmetric partner $\psi_i$ fulfill the following
boundary conditions:
\ba
gX_i(\sigma_1,\sigma_2)&=X_i(\sigma_1+2\pi ,\sigma_2)
                       &=e^{-2\pi i\theta_i}X_i(\sigma_1,\sigma_2)
                         +\pi v_i        \nonumber     \\
hX_i(\sigma_1,\sigma_2)&=X_i(\sigma_1,\sigma_2+2\pi )
                       &=e^{-2\pi i\phi_i}X_i(\sigma_1,\sigma_2)
                         +\pi u_i        \nonumber          \\
g^{-1}S_i(\sigma_1,\sigma_2)&=S_i(\sigma_1+2\pi ,\sigma_2)
                       &=e^{+2\pi i\theta_i}S_i(\sigma_1,\sigma_2)
                                         \nonumber  \\
h^{-1}S_i(\sigma_1,\sigma_2)&=S_i(\sigma_1,\sigma_2+2\pi )
                       &=e^{+2\pi i\phi_i}S_i(\sigma_1,\sigma_2) ,
\label{eq:bound}
\ea
where for an abelian orbifold $g,h$ are commuting elements of an abelian
discrete group and $u_i,v_i$ are bosonic shifts that define a lattice.
Here only fermions with diagonal boundary conditions are used, and so
a model without twists on the $H_i$ is created.
The above relations including twists {\em{may}} be fulfilled
using fermions with periodic and anti-periodic boundary conditions however.
In the case of several compactified dimensions, it has been conjectured
$Z_2,Z_4$ and $Z_8$ orbifolds are realised in this way \cite{PS89,BDL90}.
This fact also gives a simple interpretation for the connections
between the theta-functions found in Ref.\cite{PS89}.

Only for the special choice of boundary conditions above (i.e. $\theta_1$,
$\phi_1\g\in\g\{0,\frac{1}{2}\}$) do we have
the usual bosonic interpretation, by making the identification
\be
\label{ident}
\sqrt{2}\partial X \equiv
i (:{\overline\psi}_3 \psi_3:-i:{\overline\psi}_2 \psi_2:).
\ee
The algebra is then
\ba
T(z)   & = & \frac{1}{2}
            : \overline{\psi_1}\partial \psi_1:
            +: \psi_1 \partial{\overline\psi}_1:
            -
            :\partial \overline{X}\partial X : \nonumber\\
G^+(z) & = & -\sqrt{2}{\psi_1}\partial X         \nonumber\\
G^-(z) & = & -\sqrt{2}\overline{\psi_1} \partial \overline{X}
                                                \nonumber\\
J(z)   & = & :{\overline\psi_1}\psi_1:\
\label{eq:bosons}
\ea
as required. Nevertheless our approach
is in the above sense more general than that of Ref.\cite{vafa} in that
we do not require the relations in Eq.(\ref{eq:bound}) to hold
for fermionic and bosonic coordinates simultaneously.
We simply include all cases where, in the algebra in Eq.(\ref{eq:fermions}),
twist in
the fermionic $\psi_i$'s are compensated by phases in the other fermions
or after bosonisation by lattice shifts rather than twists
in the bosonic coordinates.
This corresponds to an embedding of the shifts $u_i$, $v_i$ into the
gauge group for the left movers.
We are effectively therefore considering tori with more
complicated spin-structures\footnote{Such an embedding implies a
nontrivial cohomology for the exterior derivative, $dH$, of the
field strength associated with the antisymmetric tensor $B$.
Furthermore this requires additional terms in the 10 dimensional
supergravity action used in the $\sigma$ model approach, reflecting
the non-perturbative character of the string solution.}.

It is these spin structures which can lead to more complicated models
possibly related to orbifolds. To be
more explicit, what we require for an orbifold interpretation
are identifications along the lines of Eq.(\ref{ident}), but
such identifications are of course only possible if the partition
function remains the same. This has been shown for a $Z_2$ orbifold
in for example Ref.\cite{ginsparg}. In fact searching directly for this
equivalence of partition functions was exactly the approach adopted in
Ref.\cite{PS89}.

At this point we should also state the relation to the orbifold
construction of Ref. \cite{CD88}. Here one uses the heterotic string
in a fully bosonised form for the left movers and also requires
(\ref{eq:bound}). Our interpretation belongs to using mixed $R-V$
vectors of the type
$R-V=(0,\Theta_k ,(0,v_k)\mid (\Theta_I,v_I))$, $k=1,2,3$,
$I=1,\cdots ,11$ in the authors notation.

In addition to the $Z_2$, $Z_4$, $Z_8$ orbifolds there is another
class of models that may overlap with fermionic strings. As already
pointed out in Ref.\cite{BDL90}, there is the possibility of fermionic
strings with nondiagonal boundary conditions. In the case that the
boundary conditions of the fermions transforming into each other
are the same, we find permutational orbits of fermions. Now it is
possible that such a permutational modding may be absorbed into
different diagonal boundary conditions in the fermionic string. On
the other hand it could belong to a permutational modding of the
compactified bosonic coordinates via Eq.(\ref{eq:bosonise}), or to a
phase modding in a diagonalised bosonic basis or all of the above.
This open question will be discussed further in Ref.\cite{inprogress}.

Having concluded our general discussion we now set up the framework
and the conventions for the classification.
We shall use the notation of Ref.\cite{KLT87} in which the light cone
gauge is chosen to render the string action down to that of a free
field theory on the world sheet, and all the internal degrees of freedom
are expressed as fermions. In general one expects the fermions
to transform into each other after being parallel transported
around the world sheet. However we will restrict ourselves here to
a basis in which these boundary conditions become simple phase
shifts on the fermions. In the case of
complex boundary conditions we are restricted to gauge groups of a
fixed rank of 18,20,22 in $D=8,6,4$ respectively.
A set of linearly independent `basis vectors' $\wi$
is chosen to generate all the possible boundary conditions
for the fermionic degrees of freedom in a particular model.
Sectors are represented by different points on the lattice
`spanned' by the basis vectors, $\aw = \sum_i \ai\wi$ mod(1),
where $\ai$ may take only integer values.
For the heterotic string in $D$ dimensions, the vectors $\wi$ may
take the form
\be
\wi=
\lll (s_i)^{D/2-1}\com (a^r_i\com b^r_i\com c^r_i) \ccc w^l_i \rrr,
\ee
where $r=(1,\cdots, (10-D)/2)$ and $l=(1,\cdots, 26-D)$. The
explicit form of the world sheet supercurrent also implies the triplet
constraint,
\be
a^r_i+b^r_i+c^r_i = s_i \mod .
\ee
Imposing modular invariance yields a set of vector constraints,
\ba
\label{mod1}
\kij+\kji                            & = & \wi\cdot\wj \nonumber\\
m_j \kij                             & = & 0       \nonumber\\
\kii+\kio+s_i-\frac{1}{2}\wi\cdot\wi & = & 0       \mod ,
\label{eq:modular}
\ea
where the $\kij$ are the structure constants, and $m_j$ is the least
common denominator of the boundary conditions appearing in $\wj$.
The dot product is defined by
\be
\wi\cdot\wj = \sum_{\mbox{ left }\ k} \wi^{k}\wj^{k}
            - \sum_{\mbox{ right}\ k} \wi^{k}\wj^{k}.
\ee
In addition to the above, modular invariance gives us a set of
GSO projections, which require that physical states must satisfy
\be
\label{mod}
\wi\cdot\na = \kij\aj+s_i+\koi-\wi\cdot\aw \mod ,
\ee
where $\na$ are the number operators, and a summation over repeated
indices is implied. In terms of the charge operator
\be
\label{modq}
\qa=\na+\aw-\wo,
\ee
we have
\be
\wi\cdot\qa=s_i + \kij\aj-\kio \mod.
\ee
Our purpose here is to generate $(2,2)$ models. Without loss of generality
(as we shall show presently) we choose the first four vectors to be of the
form,
\ba
\w_0 & = &
\lll (\h)^{D/2-1}\com (\h\com\h\com\h)^{(10-D)/2}
\ccc (\h\com\h\com\h)^{(10-D)/2}\com
(\h)^{(6+D)/2}\com(\h)^8 \rrr
\nonumber\\
\w_1 & = &
\lll (\h)^{D/2-1}\com (a^r_1\com b^r_1\com c^r_1)
\ccc (0\com 0\com 0)^{(10-D)/2}\com
(0)^{(6+D)/2}\com(0)^8 \rrr
\nonumber\\
\w_2 & = &
\lll (0)^{D/2-1}\com (0\com 0\com 0)^{(10-D)/2}
\ccc (a^r_1\com b^r_1\com c^r_1)\com
(\h)^{(6+D)/2}\com(0)^8 \rrr
\nonumber\\
\w_3 & = &
\lll (0)^{D/2-1}\com (0\com 0\com 0)^{(10-D)/2}
\ccc (0\com 0\com 0)^{(10-D)/2}\com
(0)^{(6+D)/2}\com(\h)^8 \rrr.
\label{eq:4vectors}
\ea
The $\wo$ vector is needed to have a modular invariant theory,
and to give the gravity multiplet.
The $\w_1$ and $\w_2$ vectors implement supersymmetry on the right and
left side (which for the heterotic string implies an $E_{D/2+4}$ gauge group).
Finally, in order to give a second $E_8^{\prime}$
factor in the gauge group we have the $\w_3$ vector. Thus we get
copies of $N=2$ algebras on each side, establishing a $(2,2)$
model. In this sense the models can be viewed as a Gepner-like construction
with internal copies of the $N=2$ minimal superconformal
models from free fields.

The numerical survey of the spectra generated by the above vectors
reveals the remarkable fact that, for any choice of
$\mbox{($a^r_1$, $b^r_1$, $c^r_1$)}$, the theory generated has the
maximal supersymmetry (and so $E_8\otimes E_8^{\prime}$ gauge groups),
and therefore corresponds to a torus compactification in the usual
sense.
Specifically this means that in $D=8,6,4$ we find $N=1,2,4$
space-time supersymmetry respectively,
and an $E_8\otimes E'_8$ gauge group.

Before continuing, we need to show that with such a choice of vectors
one may obtain all possible left-right symmetric models.
To do this we prove a useful general result, which is
that a set of $\{ \wi \}$ is equivalent to any other $\{ \zi \}$,
(i.e. gives the same set of models) provided that the boundary
conditions generated are identical. This will then ensure that the choice
of vectors in Eq.(\ref{eq:4vectors}) is sufficient to cover all
possible models without twists if one includes all values of $\kij$'s.

The proof is as follows. Since we
wish to generate the same boundary conditions we must be able
to express each $\zi $ as a linear combination of
$\{ \wi \}$. So consider the case where
we have
\ba
\zi & = &  \wi \ \  \mbox{ for }\ \ i\neq j, \nonumber\\
\zj & = & {\overline{\w_l + \wj}}, \ l\in i.
\ea
Clearly the $\zi$, $i\neq j$  projections in Eq.(\ref{mod})
for any sector $\aw$ are unchanged if $\kij^{\prime}=\kij+k_{il}$.
The $k_{ji}^{\prime}$ are then determined by Eq.(\ref{eq:modular}).
The remaining projection is
\ba
\zj\cdot\na & = & \kji^{\prime}\ai+s_j^{\prime}
                 +\koj^{\prime}-\zj\cdot\aw \nonumber\\
            & = & -(\kij+k_{il})\ai+s_j+s_l
                 +\koj+k_{0l}-\zj\cdot(\aw-\alpha\w) \nonumber\\
            & = & -(\kij+k_{il})\ai+s_j+s_l
                 +\koj+k_{0l}-(\overline{\wj+\w_l})\cdot(\aw-\alpha\w)
                                                              \nonumber\\
            & = & \wj\cdot\na + \w_l\cdot\na \mod.
\ea
Thus the modular invariance conditions are also satisfied in the new basis.
Since $\kij^{\prime}=\kij+k_{il}$ is satisfied for one choice of
structure constant, any
model generated in the old basis is also generated in the new basis.
Extrapolation to general linear combinations then follows trivially.
%
%Using this result we can also give heuristic arguments as to
%why the above gives only the maximal tori spectra, by considering
%the number of gravitinos. We first redefine $\w_1$ vectors by
%$\w_1\rightarrow\wo+\w_1$. The gravitinos are the states,
%\[
%\psi^{\mu}=\hat{O}|a\rangle \otimes {\tilde{a}}^{\mu}_{-1}|0\rangle
%\]
%where the $\hat{O}$ represents possible excitations in the compactified
%degrees of freedom, and $|a\rangle$ is the fermionic representation of the
%little group of $SO(D-1,1)$. They can therefore lie only in the
%$m_1$ sectors given by $\aw={\overline{ \alpha_1 \w_1}}$. There are
%only two modular invariance conditions of any relevance to these states;
%\ba
%\wo\cdot\qa  & = & \frac{1}{2}+\alpha_1 k_{01}-k_{00}\nonumber\\
%\w_1\cdot\qa & = & \alpha_1 k_{11}-k_{10} \mod.
%\ea
%(The ones belonging to $\bf{W_2}$, $\bf{W_3}$ are trivial in these sectors.)
%The first condition projects out half the states, and determines the
%chirality of $|a\rangle$. The second condition projects out
%$m-1$ out of every $m_1$ states.
%Since the charge lattice obeys $\frac{1}{2}\qa^2=\frac{1}{2}$
%(see Ref.\cite{KLT87}), we know that the remaining gravitino states
%must have integer masses as required. It appears that in any
%sector $\aw$ there exist two massless states for every triplet, and
%since we have $m_1$ gravitino sectors, this gives the required
%$2^{(10-D)/2}$ chiral degrees of freedom for the gravitino, after the
%$\w_1$ projection is made. (This last statement is merely an observation,
%and it was not possible to find a general proof.)

To go beyond torus compactification, we will need to add more vectors to
break down supersymmetry. Such additional vectors, which
we shall refer to as compactification vectors, may be either left--right
symmetric,
\be
\w_4 =
\lll (0)^{D/2-1}\com (a^r_4\com b^r_4\com c^r_4) \ccc
(a^r_4\com b^r_4\com c^r_4) \com
(0)^{(6+D)/2}\com(0)^8 \rrr,
\ee
or may occur in left--right symmetric pairs,
\ba
\w_4 &=&
\lll (0)^{D/2-1}\com (a^r_4\com b^r_4\com c^r_4) \ccc
(a^r_5\com b^r_5\com c^r_5) \com
(0)^{(6+D)/2}\com(0)^8 \rrr \nonumber\\
\w_5 &=&
\lll (0)^{D/2-1}\com (a^r_5\com b^r_5\com c^r_5) \ccc
(a^r_4\com b^r_4\com c^r_4) \com
(0)^{(6+D)/2}\com(0)^8 \rrr,
\label{eq:doppel}
\ea
and so on. Such a model including $\w_4$ or $\w_4$, $\w_5$
will be called "twisted", since it has the properties we already
associate with a twist in other compactifications.For example we have
an additional projection on the already existing untwisted
sectors and the appearance of additional twisted sectors.
For $N=1$ the theories generated have the gauge group
\be
\label{G}
G=g\otimes E_{(8+D)/2}\otimes E_8^{\prime},
\ee
where the first group, $g$, is some product of low rank subgroups
coming from the compactified degrees of freedom.
The selection of vectors above is not sufficient to guarantee
a (2,2) compactification since we still have to choose the structure
constants. A poor choice of $\kij$ can spoil the ($N=2$) algebra
by projecting out some of the supersymmetry generators via the modular
invariance conditions in Eq.(\ref{mod}). This can lead to
(0,2) or (2,0) models. Thus even at this stage we can have the equivalent of
Wilson line breaking.
Other compactifications like Calabi-Yau and Gepner models allow analogous
breakings. In order to guarantee a
(2,2) model we need to impose a condition on the structure constants.
We do this by insisting that, given a gauge group $G$,
there are the requisite number of gravitino degrees of freedom.

Let us now show that this is always possible by a suitable choice of
$k_{ij}$'s. The adjoints needed for building up
$G$ are always found in the $\wo$ sector. Thus we need only consider the
sectors of the form
\be
\aw =
\lll (\h)^{D/2-1}\com (\h\com\h\com \h)^{(10-D)/2} \ccc
( a^r\com  b^r\com  c^r) \com
(0)^{(6+D)/2}\com(\h)^8 \rrr.
\ee
Since the vacuum energy is $[E_R,E_L]=[-1/2,e_L]$,
they give the fermionic representation vectors for building up
the exceptional group. These are of the form
\be
\label{ferm}
A^\mu = b^\mu_{-1/2}|0\rangle\otimes \hat{O}|{a}\rangle,
\ee
where $\hat{O}$ is some combination of excitations and $|a\rangle$
is a spinorial ground state.
There always exist the sectors in which the boundary conditions of
the left and right movers are swopped
(except for the fermions building up $E'_8$), which we refer to as reflected
and denote with a tilde,
\be
\atw =
\lll (0)^{D/2-1}\com ( a^r\com  b^r\com  c^r) \ccc
(\h\com\h\com \h)^{(10-D)/2} \com
(\h)^{(6+D)/2}\com(\h)^8 \rrr.
\ee
As the vacuum energy of these sectors is
$[E_R,E_L]=[e_L,-1]$, we potentially have the gravitinos as the
reflected states of Eq.({\ref{ferm});
\be
\psi^{\mu} =\hat{\tilde{O}}|a\rangle\otimes  a^\mu_{-1}|0\rangle.
\label{eq:grav}
\ee
All that we need to do is to ensure modular invariance. We know that
the physical states in Eq.(\ref{ferm}) satisfy
\be
\wi\cdot\na = \kij\aj+s_i+\koi-\wi\cdot\aw \mod
\ee
for all $i$, so that it is sufficient to show that
\be
\twi\cdot\tna =
\tkij\taj+\tilde{s}_i+\tkoi-\twi\cdot\atw \mod.
\ee
for states of the type shown in Eq.(\ref{eq:grav}).
Since $\wi\cdot\aw+\twi\cdot\atw=0$ and
$\wi\cdot\na+\twi\cdot\nta=0$ this
translates into the condition
\be
\kij\aj+\tkij\taj+\koi+\tkoi+s_i+\tilde{s}_i=0 \mod .
\ee
In addition $\wi\cdot\wj+\twi\cdot\twj=0$ implies that
$\kij+\kji+\tkij+\tkji=0$ by Eq.(\ref{mod1}). Thus a sufficient condition
for a (2,2) compactification is
\be
\label{const}
\kij+\tkij=\delta_{j0}(s_i+\tilde{s}_i)\mod .\
\ee
So in general, for any gauge group $G$,
the appropriate gravitino states exist by a suitable choice of $\kij$.

Let us now discuss
the results of the systematic numerical study of models with
one twist, (i.e. boundary vectors  $\wo$ to $\w_4$) in $D=8$ and
$D=6$ space-time dimensions along the lines just given.

\subsection*{$D=8$ Dimensions}

The highest number of supersymmetries in $D<10$ dimensions is
$N=2^{(8-D)/2}$ and may be achieved using a torus compactification.
The models with
the highest $N$ in any dimension have no matter fields, only gauge
and gravity multiplets. Thus in $D=8$, since there is only $N=1$
supersymmetry, singlets together with gauge
bosons and their superpartners form full gauge supermultiplets,
and the compactification is essentially trivial.

Nevertheless it will
prove useful to examine this case, since it will allow us to observe
some general aspects of our scheme which we can apply to the more
complicated lower dimensional models.
The most obvious feature of  (2,2) compactifications is the dramatic
reduction in the number of models. In fact searching over the
$\sim 10^5$ models up to order 20 (which very probably
contain all possible distinct models),
we find only two supersymmetric ones.
These models, which have $E_8\otimes E_8^{\prime}$ symmetry are shown in
table (1).

In addition we find that the different models are
generated by
the compactification vectors, regardless of the supersymmetry vectors.
Thus given the order of the supersymmetry, a model (i.e. the gauge
group, number of generations and singlets) is defined mainly by the
compactification vectors.

Let us now compare the results of table (1)
with other compactifications. Because of the requirement of absence
of tachyons, there exist no orbifolds.
{}From the table it is clear that there is no equivalent to the
maximal Gepner model with $g=SU(3)$ \cite{G88}, but an additional
model with $g=SU(2)\otimes U(1)$. In fact the first
configuration for an $SU(3)$would require the boundary conditions
of the internal
degrees of freedom to be degenerate. This is disallowed by the
modular invariance conditions. Despite this it does seem that fermionic
strings and Gepner models in $D=8$ correspond to the torus
at different points in moduli space. However a direct bosonisation
as discussed in the introduction is only possible with periodic and
anti-periodic boundary conditions resulting in model 1. Here one also
realises that no orbifolds are possible due to the modular invariance
conditions.

Alternatively the fermionic strings
could also be interpreted as generalised tori as outlined above.
Furthermore, we find that the supersymmetry may be broken entirely by
adding just one $\w_4$, if the phases chosen are complicated enough.
This contrasts with the techniques used by most model builders who
frequently introduce many vectors with the simplest twistings.

\subsection*{$D=6$ Dimensions}

In this case we may have $N=2$ with $G=g\otimes E_8\otimes E_8^{\prime}$,
or $N=1$ with $G=g\otimes E_7\otimes E_8^{\prime}$.
We searched through $\sim 10^6$ models and find only 37 distinct cases.
These are displayed in table (2).

As in the case of $D=8$, we find that imposing $(2,2)$ symmetry
drastically reduces the number of available models.
The models are repeated many times with vectors of arbitrarily high order.
The minimal allowed gauge group is $U(1)^5$, and the number of generations
is nearly always less than ten which corresponds to the K3 manifold and the
$Z_N$ orbifolds. We observe that, with the exception of three models,
the number of singlets
is a multiple of the number of generations.

Clearly for the first six models the spectrum of the
the $T^2_2$ torus with enlarged gauge group emerges.
Bosonisation is restricted as above and gives models 1 and 3.
In the fully bosonised
formalism simple torus compactification corresponds to having only one
vector $V_0$ (in which all the entries are $\h$ in the notation
of Ref.\cite{CD88}) and gives $N=1,2,4$ and $E_8\otimes E_8^{\prime}$ in
$D=8,6,4$ dimensions respectively). In the case of the Gepner models
one obtains $g=SU(3)\ox SU(2)^2$.

However, at this point we are also able to identify four spectra
in models 9--12 in which the number of generations, and of
untwisted generations match those of the four $Z_N$ orbifolds in
$D=6$ \cite{HM87}. These orbifolds have already been studied
in Ref.\cite{HM87} and it was found that all of them may be blown up into
the $K3$ manifold. For example, the choice of vectors which have the
non-zero entries
\be
c_1^r=\h,\g a_4^r=\frac{1}{6},\g
b_4^r=c_4^r=\frac{11}{12}\g\g\g (r=1,2)
\ee
generates model 10. In addition to the single untwisted generation,
we find single generations coming from the
sectors $\overline{\wo\pm 2\w_4}$,
 $\overline{\wo\pm 4\w_4}$ and
 $\overline{\w_1+\w_2\pm 2\w_4}$,
and 3 generations from each of the sectors
$\overline{\wo+6\w_4}$ and $\overline{\w_1+\w_2+6\w_4}$. (These sectors
give the $\underline{12}$ representation and acting with the analog of the
supersymmetry on the left side ($\w_2$) gives the $\underline{32}$ to
build up $\underline{56}$ of $E_7$.) The generations in model 10 are
always distributed in this way. In fact there is a particular
characteristic distribution of generations for each of the models 9--12.

This is an intriguing connection,
but direct reformulation as an orbifold along the lines given in the
introduction is only possible for the case of the $Z_2$ orbifold. This is
acheived using vectors which have only periodic or anti-periodic
boundary conditions. There are only two possible non-trivial cases.
The first has a single $\w_4$ vector with the entries
\be
c_1^r=\h,\g b_4^r=c_4^r=\h \g\g\g (r=1,2).
\ee
The second can be made from the first by adding an additional
symmetric vector, $\w_5$, to the above, with entries
\be
a_5^1=c_5^1=b_5^2=c_5^2=\h
\ee
and all others zero. Both models have exactly the same spectrum
as the $Z_2$ orbifold \cite{HM87} accompanied by additional, matching
pairs of gauge bosons and singlets. Thus we see that the relation
of these fermionic string models to the $Z_2$ orbifold is identical to
that between the superconformal models and the K3 manifold. We therefore
deduce that the fermionic string belongs to a $Z_2$ orbifold
on a point in moduli space with enlarged symmetries. Using real
fermions one may then break the rank of the gauge group. We should
stress that in $D=6$ the modular invariance conditions prohibit the
construction of the $Z_4$ orbifold in a similar manner even with
real fermions.

In addition to the above, we have two rather peculiar models with 13 and
17 generations. These have particularly symmetric configurations. For
example the series of models with the non-zero boundary conditions
\be
c_1^r=\h,\g b_4^1=\frac{1}{m},\g c_4^1=c_4^2=\frac{m-1}{m},
\g a_4^2=\frac{m+1}{3m},\g b_4^2=\frac{2(m+1)}{3m}
\g\g\g m=2^{(2n+1)}
\ee
generates only these spectra.

Having made a systematic study of models with one compactification vector
of the $\w_4$ type, let us make some remarks about the case of
symmetric pairs (\ref{eq:doppel}). If we choose a $\w_4$, $\w_5$ with
$(a_5^r,b_5^r,c_5^r)=(0,0,0)$,
then it is possible to generate more vectors of the $\w_1$ variety.
In this case the $\w_4$ and $\w_5$ vectors usually
project out as many gravitino degrees of freedom as new
ones are generated, and the order of the supersymmetry
is unchanged. The gravitinos may then appear in more complicated sectors
of the form $\aw=\overline{\wo+\alpha_1\w_1+\alpha_4\w_4}$, which
depend on the choice of the structure constants $\kij$.
We have found this to be true for various cases and we believe this
to be a general mechanism.

For the general case of several compactification vectors no simple
pattern is obvious, but the classification of models so far suggests
that there should be only models with a smaller number of generations
than the those already found.

The results of our classification of fermionic strings in $D=8$
and $D=6$ dimensions are the following. In $D=8$ we find only the
torus with different gauge groups $g$. In $D=6$ we find nearly all the
possible generation numbers below 10, which is the result for the
$K3$ manifold and $Z_N$ orbifolds, and in addition models with
13 and 17 generations. The overlap with orbifolds consists of the $Z_2$
compactification, for which a direct bosonisation procedure exists
even in this case of purely complex fermions. The spectrum of
the $Z_2$ orbifold is generated in many more cases where direct bosonisation
is not possible, and this leads us to conjecture that the
three additional 10 generation models may be linked to the
$Z_3$, $Z_4$ and $Z_6$ orbifolds.
This will possibly be explained by a study of fermionic strings
with permutational moddings. Either way it seems to be always possible to
interpret fermionic strings as tori with generalised spin structures.

A similar study for $D=4$ is under way and should shed more light
on these questions. There a systematic study is much more
complicated because of the huge number of possible models.
Nevertheless our observations in higher dimensions have given some idea
as to what we can expect to find.

\vspace{1cm}
\noindent
{\bf\Large Acknowledgement} \hspace{0.3cm} We thank D.~C.~Dunbar and
H-P.~Nilles for useful discussions. We would also like to thank the RAL
computer division, especially Dick Roberts.

\bigskip
\bigskip

\section*{Table Captions}

\begin{description}

\item{\bf Table 1 }Supersymmetric (2,2) models in $D=8$.
$U$ is the number of untwisted generations, and $n_s$ is
the number of singlets. The gauge group is
$g\otimes E_8\otimes E^{\prime}_8$.
The singlets together with the gauge
bosons and their superpartners form full gauge supermultiplets.
Where possible we express $g$ as a product of special unitary groups.

\item{\bf Table 2 }Supersymmetric (2,2) models in $D=6$.
The gauge group is $g\otimes E_{6+N}\otimes E^{\prime}_8$.
For $N=2$ the singlets are incorporated in gauge multiplets.

\end{description}

\clearpage

\begin{table}
\caption{\hspace{16 cm}}
\begin{tabular}{|l|l|l|l|l|l|l|}\hline\hline
 number &$N$&        group $g$      & gen&$\gen$ &$U$& $n_s$ \\ \hline
     &   &  \mbox{                 }  &    &    &   &     \\
  1  & 1 & $SU(2)^2$                  &  2 & 2  & 4 & 6   \\
  1  & 1 & $SU(2)\otimes U(1)$        &  2 & 2  & 4 & 4   \\
\hline\hline
\end{tabular}
\end{table}

\mbox{}

\clearpage

\begin{table}
\caption{\hspace{16 cm}}
\begin{tabular}{|l|l|l|l|l|l|l|}\hline\hline
 number &$N$&        group $g$      & gen&$\gen$ &$U$& $n_s$ \\ \hline
     &   &                          &    &    &   &     \\
  1  & 2 & $SO(8)$                  &  2 & 2  & 4 & 34  \\
  2  & 2 & $SU(4)\ox U(1)$          &  2 & 2  & 4 & 22  \\
  3  & 2 & $SU(2)^4$                &  2 & 2  & 4 & 18  \\
  4  & 2 & $SU(2)^3\ox U(1)$        &  2 & 2  & 4 & 16  \\
  5  & 2 & $SU(2)^3\ox U(1)$        &  2 & 2  & 4 & 16  \\
  6  & 2 & $SU(2)^2\ox U(1)^2$      &  2 & 2  & 4 & 14  \\
  7  & 1 & $U(1)^5$                 &  17& 0  & 1 & 51  \\
  8  & 1 & $U(1)^5$                 &  13& 0  & 1 & 52  \\
  9  & 1 & $SU(2)^5$                &  10& 0  & 2 & 80  \\
  10 & 1 & $SU(2)^2\ox U(1)^3$      &  10& 0  & 1 & 32  \\
  11 & 1 & $SU(2)\ox U(1)^4$        &  10& 0  & 1 & 54  \\
  12 & 1 & $U(1)^5$                 &  10& 0  & 1 & 30  \\
  13 & 1 & $U(1)^5$                 &  9 & 0  & 1 & 36  \\
  14 & 1 & $U(1)^5$                 &  9 & 0  & 1 & 27  \\
  15 & 1 & $SU(2)^3\ox U(1)^2$      &  8 & 0  & 2 & 40  \\
  16 & 1 & $SU(2)^2\ox U(1)^3$      &  7 & 0  & 1 & 20  \\
  17 & 1 & $SU(2)\ox U(1)^4$        &  6 & 0  & 2 & 30  \\
  18 & 1 & $U(1)^5$                 &  5 & 0  & 1 & 25  \\
  19 & 1 & $U(1)^5$                 &  5 & 0  & 1 & 20  \\
  20 & 1 & $U(1)^5$                 &  5 & 0  & 1 & 15  \\
  21 & 1 & $SU(4)\ox U(1)^2$        &  2 & 0  & 1 & 14  \\
  22 & 1 & $SU(4)\ox U(1)^2$        &  2 & 0  & 1 & 12  \\
  23 & 1 & $SU(2)^2\ox U(1)^3$      &  2 & 0  & 1 & 10  \\
  24 & 1 & $SU(2)^2\ox U(1)^3$      &  2 & 0  & 1 & 8   \\
  25 & 1 & $SU(2)^2\ox U(1)^3$      &  2 & 0  & 2 & 6   \\
  26 & 1 & $SU(2)\ox U(1)^4$        &  2 & 0  & 1 & 6   \\
  27 & 1 & $SU(4)\ox U(1)^2$        &  1 & 0  & 1 & 6   \\
  28 & 1 & $SU(3)\ox U(1)^3$        &  1 & 0  & 1 & 10  \\
  29 & 1 & $SU(2)^3\ox U(1)^2$      &  1 & 0  & 1 & 10  \\
  30 & 1 & $SU(2)^3\ox U(1)^2$      &  1 & 0  & 1 & 8   \\
  31 & 1 & $SU(2)^2\ox U(1)^3$      &  1 & 0  & 1 & 5   \\
  32 & 1 & $SU(2)^2\ox U(1)^3$      &  1 & 0  & 1 & 4   \\
  33 & 1 & $SU(2)\ox U(1)^4$        &  1 & 0  & 1 & 6   \\
  34 & 1 & $SU(2)\ox U(1)^4$        &  1 & 0  & 1 & 5   \\
  35 & 1 & $SU(2)\ox U(1)^4$        &  1 & 0  & 1 & 3   \\
  36 & 1 & $U(1)^5$                 &  1 & 0  & 1 & 4   \\
  37 & 1 & $U(1)^5$                 &  1 & 0  & 1 & 3   \\
\hline\hline
\end{tabular}
\end{table}

\end{document}